\documentclass[a4paper]{article}

\usepackage{atmohead2013}
\usepackage[english]{babel}
\usepackage{epstopdf} 

\title{Atmospheric considerations for the CTA site search}

\authors{
St\'{e}phane Vincent$^{1}$
for the CTA Consortium.
}

\afiliations{
$^1$ DESY, Platanenallee 6, 15738 Zeuthen, Germany \\
}

\email{stephane.vincent@desy.de}

\abstract{
The Cherenkov Telescope Array (CTA) will be the next high-energy gamma-ray observatory. 
Selection of the sites, one in each hemisphere, is not obvious since several factors have 
to be taken into account. 
Among them, and probably the most crucial, are the atmospheric conditions. 
Since July 2012, the site working group has deployed automatic ground based instrumentation (ATMOSCOPE) 
on all the candidate sites. 
Due to the limited time span available from ground based data, long term weather forecast models become 
necessary tools for site characterization. 
It is then of prime importance to validate the models by comparing it to the ATMOSCOPE measurements. 
We will describe the sources of data (ATMOSCOPE, weather forecasting model and satellite data) for the 
site evaluation and how they will be used and combined.
}


\begin{document}
\maketitle

\section{Introduction}
CTA will consist of two arrays of Cherenkov telescopes\cite{ach13}, which aim to: 
(a) increase sensitivity by an order of magnitude with respect to the existing observatories like H.E.S.S., 
MAGIC and VERITAS, 
(b) increase the detection area and hence detection rates,
(c) improve the angular resolution and hence the morphology reconstruction of extended sources,
and (d) provide energy coverage for photons from some tens of GeV to beyond 100 TeV. 
The observatory will consist of two arrays: a southern hemisphere array, which allows a deep 
investigation of galactic sources and of the central part of our Galaxy, but also for the 
observation of extragalactic objects. 
The northern hemisphere array is primarily devoted to the study of Active Galactic Nuclei and 
galaxies at cosmological distances. 
The arrays will also make contributions to the field of particle physics with searches for dark matter, 
tests of Lorentz invariance and searches of axion-like particles. 
CTA will be operated as an open, proposal-driven facility.

In order to find the best location for the construction of CTA, a comprehensive site search campaign 
was conducted. 
The criteria considered for the site ranking are science performance, which depends on the average 
annual observation time available and the performance of the array per unit time at a given site, as well 
as costs and risks. 
Requirements which are most relevant for the site selection are:
\begin{itemize}
\item Altitude of the sites between 1500 - 3800m above sea level.
\item Area available for the deployment must be $>$10km$^2$ for the South and $>$1km$^2$ for the North.
\item Ground slope must be less than 8\%.
\item The site should have $>70\%$ of moonless night hours completely cloud free.
\item During observations the 10-minutes average wind speed is $<36$ km.h$^{-1}$.
\item Ambient air temperature during observations will be -15 to +25$^{\circ}$C.
\end{itemize}
Following the call for site proposals, a total of nine site proposal were received. 
Five possible locations in the southern hemisphere are evaluated: 
two locations in Argentina, two in Namibia and one in Chile. 
The northern proposals include a site in Mexico, two sites in the USA and a site on 
the island of Tenerife in Spain. 
The locations of the nine site sites taken into considerations are given in Table  \ref{candidate_site}. 
The Argentinian sites are described in \cite{all13}, the U.S. sites are presented by \cite{ong13}, and for the 
Spanish site see \cite{pue13}.

\section{Data sources}
The best way to compare the atmospheric conditions at the different sites is to have long term data from 
ground instruments and observations on the actual sites. 
The CTA consortium has deployed ground based instrumentation on all sites, 
has analyzed satellite data and obtained retrodictions of environmental conditions from weather simulations. 
Correlation studies with simultaneous data have been performed and the results from these studies give us 
a picture of the long-term weather patterns at the CTA candidate sites. 
Some sites are already extensively characterized and hence serve rather for the validation of the 
instruments and methods. 

\begin{center}
\begin{table*}[ht]
{\small
\hfill{}
\begin{tabular}{|l|l|c|c|c|c|c|c|c|}
\hline
Name & Country & Latitude[$^{\circ}$] & Longitude[$^{\circ}$] & Elevation[m]\\
\hline
Leoncito  & Argentina  & 31.7S & 69.3W & 2600 \\ \hline
San Antonio de los Cobres & Argentina & 24.0S & 66.2W &3600 \\ \hline
Armazones & Chile & 24.6S & 70.2W & 2400 \\ \hline
H.E.S.S. & Namibia & 23.2S & 16.5E & 1850 \\ \hline
Aar & Namibia & 26.7S & 16.4E & 1650 \\ \hline
Tenerife & Spain & 28.3N & 16.5W & 2200 \\ \hline
Meteor Crater & USA & 35.0N & 111.0W & 1700 \\ \hline
Yavapai & USA & 35.1N & 112.9W & 1650 \\ \hline
San Pedro Martir & Mexico & 31.0N & 115.5W & 2500 \\ \hline
\end{tabular}}
\hfill{}
\caption{List of the sites under considerations for the construction of CTA.}
\label{candidate_site}
\end{table*}
\end{center}

\subsection{Ground based instrumentation}
\begin{figure}[!h]
 \centering
 \includegraphics[width=0.4\textwidth]{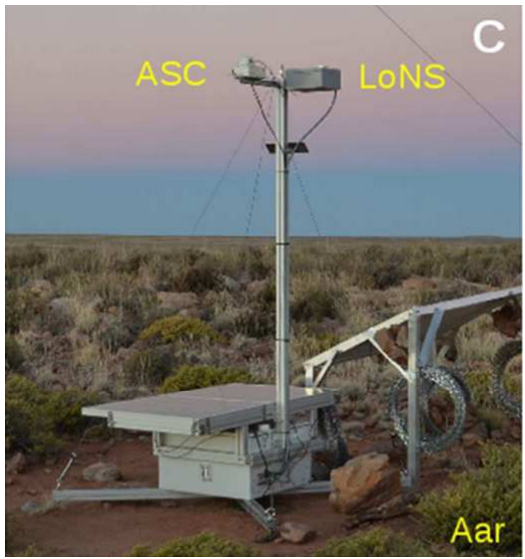}
 \caption{The ATMOSCOPE at Aar.}
 \label{atmoscope}
 \end{figure}

The ATMOSCOPE (Autonomous Tool for Measuring Site COndition PrEcisely) 
is a sensor station made to be deployed and operated in a remote site without power supply and 
ethernet connection. 
The instrumentation consists of: 
\begin{itemize}
\item A commercial weather station capable of measuring temperature, humidity, pressure and wind characteristics. 
A pyrometer is used as a cloud altitude sensor. 
\item A Light of Night Sky (LoNS) instrument measuring the light of the night sky through a B band filter and through a 
V band filter.
\item An All Sky Camera (ASC) was developed as an device for the monitoring of the night sky quality. 
The analysis results are the cloud fraction and night sky brightness.
\end{itemize}
Further details on the LoNS can be found in \cite{gau13} and details of the ASC can be found in \cite{man13}. 
The calibration of the ATMOSCOPE was initially performed after their construction. 
While at sites, the calibration of the LoNS is regularly performed with a standard light source for several light levels. 
All the ATMOSCOPE stations are connected to the internet and at the remote sites this has been achieved by using GPRS connection through local cell phone networks. 
Figure \ref{atmoscope} gives an image of an ATMOSCOPE on location at a candidate site. 
In addition to the instruments, mirror samples are located on the structure to evaluate the degradation of the 
reflective surfaces due to erosion by wind-blown dust.
The amount of data gathered by the instruments varies from site to site and the aim is to gather at least one year 
of data from each site.

\subsection{Weather simulations}
In order to overcome the limited time span available from ground based measurements, several contracts have been placed 
with an external company (SENES, canada). 
The weather simulations provide weather parameters with a spatial precision of 1km$\times$1km 
and with one sets of data parameter per hour for a period of 10 years or more. 
The data parameters include: cloud cover; wind; temperature; humidity and precipitation. 
The precipitation data give the amount of snow and hail as well as information on snow cover on the ground. 
Further the simulation gives estimates of dust deposition

The SENES weather simulation is a state-of-the-science weather forecast model using historical data to develop 
climatology of clear nights and adverse environmental conditions for the candidate locations. 
The code "FReSH-4" is a SENES-developed weather forecasting that includes 3 modules. 
The first module collects and formats the global base temperature and pressure observational data. 
The second module runs an operational forecast model \cite{wrf}. 
The third module tailors the outputs from the model. 
The dust predictions come from the Dust REgional Modeling (DREAM) system. 
This model uses the wind speed and direction information predicted by the ETA weather model 
\cite{eds} to determine when the wind speed at the surface is high enough to re-suspend dust particles into 
the atmosphere. 

 \begin{figure}[!h]
 \centering
 \includegraphics[width=0.4\textwidth]{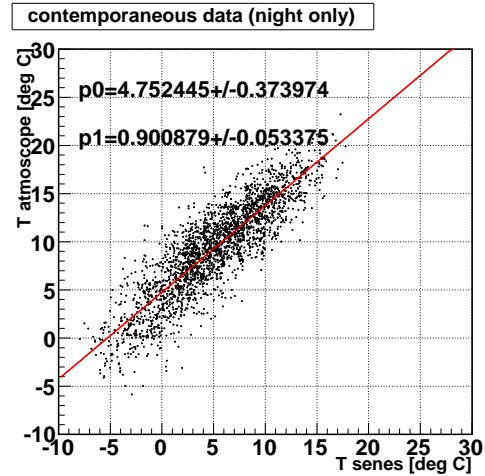}
 \caption{Example of controls of SENES weather simulation data by comparing measurements from 
 ground based weather stations with the simultaneous SENES data.}
 \label{atmoscope_senes_comparison}
 \end{figure}
\subsection{Comparison of ground based data and weather simulations for environmental quantities}
For the weather data: wind, temperature and relative humidity, the SENES simulation data could, in principle, 
be used to provide the loss term for bad weather condition and would have the advantage of long time periods 
compared to the ATMOSCOPE information. 
An example of the comparison of the SENES simulation and ground based weather station is given in 
Figure \ref{atmoscope_senes_comparison}. 
This example shows a reasonable correlation between the measurements and the simulations. 
Extensive studies have been made to assess the reliability of these data \cite{vin13}.

\subsection{Satellite imaging for cloud coverage measurement}
As a source of data for the determination of cloud fraction, we are using data from several satellites. 
This includes data from three different satellite families; two of them are geostationary 
(European METEOSAT and U.S. GOES), one is polar (MODIS). 
These satellites have very extensive archives of data, spanning typically over a decade or more. 

\begin{figure}[!h]
 \centering
 \includegraphics[width=0.4\textwidth]{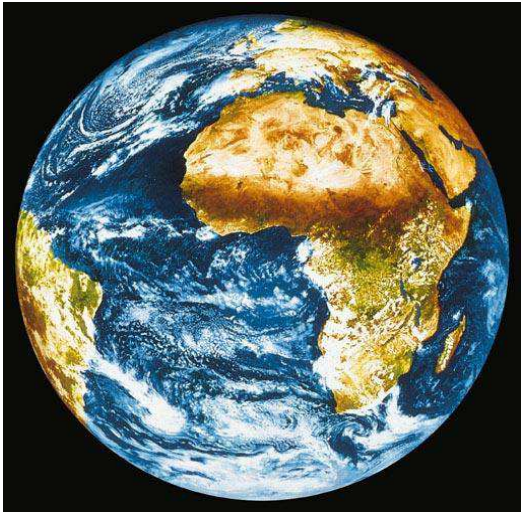}
 \caption{Earth from the geostationary weather satellite METEOSAT.}
 \label{meteosat_fov}
 \end{figure}

In general, polar satellites are on orbit much closer to the Earth, therefore they have typically much 
better resolution. 
They rotate quickly around the Earth and are able to view every candidate sites typically once per day. 
The geostationary satellites are much further away from the Earth, their imagers have thus poorer resolution. 
On the other hand, the imagers take typically one image per 15 minutes. 
As the sites in general do not have the same amounts of cloud cover during the night and day, the good 
time resolution is therefore essential for proper determination of the average cloudiness above sites. 
The field of view of the METEOSAT satellite contains the Canary Islands, the Namibian sites 
and parts of South America (close to the edge of field of view).  
Five of the candidate sites are covered by this satellite: Aar, H.E.S.S., Tenerife, Leoncito and San Antonio de 
los Cobres (see Figure \ref{meteosat_fov}). 

\begin{figure}[!h]
 \centering
 \includegraphics[width=0.5\textwidth]{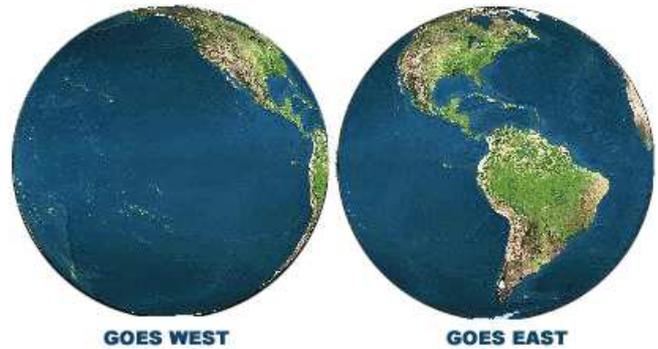}
 \caption{Field of view for the GOES satellites.}
 \label{goes_fov}
 \end{figure}

GOES data cover the American sites and the Canary Islands, which are at the edge of the field of view 
(see Figure \ref{goes_fov}). 
The satellite studies have been found to be more precise and reliable for flat areas than for sites with 
abrupt orography. 

Finally, the third dataset used for comparison comes from the MODIS satellites which cover all 9 
candidate sites. 
One or two data points are available per day. 
The data were selected to contain only measured values during nigh time and the clear fraction 
averages were computed. 

\subsection{Other data source and site visits}
Further data sources have been searched for. 
These sources include ground based meteorological data provided by National Agencies. 
Other sources have been found from miscellaneous scientific publications and websites. 

\section{Summary}
CTA will be the next generation of imaging Cherenkov telescopes with  
one observatory in the Northern hemisphere and one in the Southern hemisphere. 
The activities to search for sites for CTA started in 2008. 
Following a call for site proposals, a total of nine proposal documents were received. 
 
Here we have described the basic ideas used for the site search.  
The site characterization depends on 
three main sources of information. 
For most sites, a year of ground based data (ATMOSCOPE data) will be gathered. 
As this is not sufficient to make a comprehensive comparison of the sites, we must use several long 
term data sources from sites, such as the numerical weather simulations and the satellite data. 
A major challenge in evaluating the data is to understand the possible biases and uncertainties. 
In order to overcome these difficulties, a comparison between the ground based data and the long term 
data sources is being conducted. 
This allows to find uncertainties and systematic shifts in the data. 
The data are also verified for consistency at other nearby sites: at local astronomical observatories and weather 
stations. 
In most cases, the comparison between the ground based and other sources of data is satisfactory, yet there are some 
cases when we can only rely on the limited ground based data. 

There are also several site related studies carried by the CTA consortium. 
The potential influence of the presence of the E-ELT lasers in Chile is shown in \cite{gau13b}. 
There is a mirror test facility in San Antonio de los Cobres in Argentina and is described in \cite{med13}.  
The CTA site decision will also take into account the estimate of the infrastructure cost. 
The decision process is currently under way and we expect to have a final recommendation at the end of 2013.

\vspace*{0.5cm}
\footnotesize{{\bf Acknowledgment: }{
We gratefully acknowledge support from the agencies and organizations 
listed in this page: http://www.cta-observatory.org/?q=node/22.
SV acknowledges support through the Helmholtz Alliance for Astroparticle Particle. 
}}

\end{document}